\documentclass[12pt,preprint]{aastex}
\def\lcdm{$\Lambda {\rm CDM} \,\,$} 
\def\b{\begin{equation}}
\def\e{\end{equation}}

\begin{document}
\title{Using the filaments in the LCRS to test the $\Lambda$CDM model}
\shorttitle{Using  filaments in  LCRS to test $\Lambda$CDM}
\author{Somnath Bharadwaj and Biswajit Pandey}
\affil{Department of Physics and Meteorology \& Center for
  Theoretical Studies \\ Indian Institute of Technology, Kharagpur 721
  302, India} 
\email{somnath@cts.iitkgp.ernet.in, pandey@cts.iitkgp.ernet.in}

\begin{abstract}
It has recently been established that the filaments seen in the Las
Campanas Redshift Survey (LCRS) are statistically significant at 
scales as large as $70$ to $80 \, h^{-1} {\rm Mpc}$ in the
$\delta=-3^{\circ}$ slice,  and $50$ to $70 \, h^{-1} {\rm Mpc}$ in the
five other LCRS slices. The ability to produce such filamentary
features  is an important test of any model for structure
formation. We have tested the \lcdm  model with a
featureless, scale invariant primordial power spectrum by quantitatively
comparing the filamentarity in simulated LCRS slices with the actual 
data. The filamentarity in an  unbiased \lcdm model, we find,  
is less than the LCRS. Introducing  a bias $b=1.15$,  the model is in 
rough 
consistency  with the data, though in two of the  slices the
filamentarity  falls below the data  at a low level of statistical
significance. The filamentarity is  very sensitive to the  bias
parameter and a high value  $(b=1.5)$, which enhances filamentarity at
small scales and suppresses it at large scales, is ruled out. A bump
in the power spectrum at $k \sim 0.05 \, h \, {\rm Mpc^{-1}}$ is found to
have  no noticeable effect on the filamentarity.   
\end{abstract}

\section{Introduction}
Quantifying the clustering pattern observed in the galaxy distribution 
and explaining its origin has been one of the central themes 
in modern cosmology (eg. Peebles 1980). Traditionally, correlation
functions  have been used for this purpose, with the  two-point
correlation function  $\xi(r)$ and its Fourier transform, the power
spectrum $P(k)$, receiving most of the attention. There now exist
precise estimates of $\xi(r)$  (eg. Tucker et al. 1997, LCRS; Hawkins  
et al. 2003, 2dFGRS; Zehavi et al. 2002  SDSS) and  $P(k)$
(eg. Lin et al. 1996, LCRS  ; Percival et al. 2001, 2dFGRS; Tegmark
et al. 2003a, SDSS ) determined from several extensive redshift
surveys. It is found that the large scale clustering of galaxies is
well described by a featureless, adiabatic,  scale invariant
primordial power spectrum in a $\Lambda {\rm CDM}$ cosmological
model. We shall, henceforth, refer to this combination of the
background cosmological model and the power spectrum as the \lcdm
model. The observed  CMBR anisotropies  (eg. Spergel et al. 2003,
WMAP),  and the joint analysis of  CMBR anisotropies and galaxy
clustering data  (eg. Tegmark et al., 2003b),  which place very precise
constraints on cosmological models, are all  consistent with the \lcdm
model. This model is now generally accepted as the minimal model which
is consistent with most currently available cosmological data.

One of the most striking visual features in all the galaxy redshift
surveys  e.g., CfA (Geller \& Huchra 1989), LCRS (Shectman et
al. 1996),  2dFGRS (Colless et al. 2001, Colless al. 2003) and SDSS
(EDR) (Stoughton, et al. 2002, Abazajian et al. 2003).  
is that the galaxies  appear to be distributed along filaments.
These filaments are interconnected and form a network known as 
the ``Cosmic Web''. The region in between the filaments are voids
which are largely devoid of galaxies. This description of the galaxy
distribution, based on the morphology of coherent structures observed
in redshift surveys, addresses an aspect of the clustering pattern
which is completely missed out by the two-point statistics like
$\xi(r)$ and $P(k)$. In the currently accepted scenario of structure
formation, the primordial density perturbations are a Gaussian random
field which is completely described by the power spectrum, the phases
of different modes being random. As structure formation progresses,
coherent structures like sheets and filaments are formed
through the process of gravitational instability. The phases of
different Fourier modes are now correlated,  and the power spectrum
does not fully describe the statistical properties of the large scale 
structures. The full hierarchy of N-point statistics can,  in
principle, be used to quantify all  properties of the large-scale
structures, but these, being hard to determine at large scales,  have
not been perceived as the optimum tool for this purpose. Direct
methods of 
quantifying the morphology and topology of large scale coherent
features (as we discuss later) have been found to be more effective. 
 The ability to produce large-scale coherent structures
like the filaments observed in  galaxy surveys is an important
test of any  models of structure formation. Any such test addresses
issues beyond the scope of the two-point statistics like the power
spectrum.  In this paper we ask the question whether the
\lcdm model is consistent with the filaments  observed  in
galaxy redshift surveys.  

The analysis of filamentary patterns in the galaxy distribution has a
long history dating back to papers by Zel'dovich, Einasto and Shandarin
(1982), Shandarin and Zel'dovich (1983) and Einasto et al. (1984) . In
the last paper the authors analyze the distribution of galaxies in the
Local Supercluster where they find filaments whose lengths increase
with smoothing and finally  get interconnected into an infinite
network of superclusters and voids. The percolation analysis and the
genus statistics (Zel'dovich et al. 1982; Shandarin \& Zel'dovich
1983, Gott, Dickinson \& Mellot 1986) were some of the earliest
statistics introduced to 
quantify the network like topology of the galaxy distribution. 
 A later study (Shandarin \& Yess 2000) used percolation analysis to
show the presence of a network like structure in the distribution of
the LCRS galaxies. The large-scale and super 
large-scale structures in the distribution of the LCRS galaxies have
also been studied by Doroshkevich et al. (2001) and Doroshkevich et
al. (1996) who find evidence for a network of sheet like structures
which surround underdense regions (voids) and are criss-crossed by
filaments.  The distribution of voids in the LCRS has been studied by
M{\" u}ller, Arbabi-Bidgoli, Einasto, \& Tucker (2000) and the
topology of the LCRS by Trac, Mitsouras, Hickson, \&
Brandenberger (2002) and Colley (1997).  A recent analysis (Einasto et
al. 2003) indicates a super cluster-void network in the Sloan Digital
Sky Survey also. Colombi, Pogosyan and Souradeep (2001) have studied
the topology of excursion sets at the percolation threshold.  The
minimal spanning tree (Barrow,Bhavsar \& Sonoda  1985) is another
useful way to probe the geometry of large scale structures. The
morphology of  superclusters in the PSCz has been studied by  
Basilakos,  Plionis \&  Rowan-Robinson (2001) who find filamentarity 
to be the dominant feature. On comparing their results with
the predictions of different cosmological models, they find that a low
density  $\Lambda$CDM model is preferred. These conclusions were
further confirmed by Kolokotronis, Basilakos \&  Plionis (2002) using
the Abell/ACO cluster catalogue.  

The Minkowski functionals have  been suggested as a novel tool to
study the morphology  of structures in the universe (Mecke et
al. 1994; Schmalzing \& Buchert 1997). Ratios of the Minkowski
functionals can be used to define  
a shape diagnostic 'Shapefinders'  which faithfully quantifies  the
shapes of both simple and topologically complex objects  (Sahni,
Satyaprakash \& Shandarin 1998).   Bharadwaj et al. (2000) 
defined the Shapefinder statistics in two dimensions (2D),  and  used
this  
to demonstrate that the galaxy distribution in the LCRS
exhibits a high degree of filamentarity compared to a random Poisson
distribution having the same geometry and selection effects as the
survey. This analysis provides objective confirmation of the visual 
impression that the galaxies  are distributed along filaments.
In a later paper Bharadwaj, Bhavsar and Sheth (2004) used Shapefinders
in conjunction with a statistical technique  called Shuffle (Bhavsar
\& Ling 1988) to determine the maximum length-scale at  which the
filaments observed in the LCRS are statistically significant.
They found that the largest length-scale at which filaments are
statistically significant is between  $70$ to $80 \,
h^{-1}$Mpc, for the LCRS $-3^o$ slice. Filamentary 
features longer than $80 \, h^{-1}$Mpc, though identified, are not
statistically significant. Such features arise from chance
alignments of galaxies. Further,  for the five other LCRS slices, filaments of
lengths $50 \, h^{-1}$Mpc to  $70 \, h^{-1}$Mpc were found to be
statistically significant, but not beyond. 

Comparing the filamentarity observed in galaxy redshift surveys
against the predictions of different models of structure formation 
provides a unique method for testing these models. Here we present a
method for carrying out this analysis in 2D, and as an example we
apply it to the LCRS for which the filamentarity has already been
extensively studied ( Bharadwaj et al. 2000; Bharadwaj et al. 2004). 
In this paper we address the question if the \lcdm  model is
consistent with the filaments observed in the  LCRS. We have used
cosmological N-body simulations to generate different realizations of
the galaxy distribution one would expect in the LCRS  for the $\Lambda
{\rm CDM}$ model. The  actual and simulated LCRS data were  analyzed
in  exactly the same  way using  Shapefinders to quantify the degree
of filamentarity,  and the results were compared to test  if the
predictions of the  $\Lambda{\rm CDM}$ model are consistent with  the
LCRS. 

The LCRS galaxies may be a biased tracer  of the underlying dark
matter distribution whose evolution is followed by the N-body
simulation. We also consider this possibility,  and study how varying
the bias parameter effects the network of filaments and voids. 

Various independent lines of investigation seem to indicate 
that there may be excess power, in comparison to the \lcdm
model, at scales $k \sim 0.05 \, h \, {\rm Mpc}^{-1}$. The two
dimensional power spectrum for the LCRS (Landy et al. 1996) exhibits
strong excess power at wavelengths $\sim 100 \, h^{-1}$Mpc. 
The analysis of the distribution of Abell clusters (Einasto et
al. 1997a, 1997b) reveals  a bump in the power spectrum 
at  $k=0.05 \, h \, {\rm Mpc}^{-1}$.  Also, the recent analysis of the
SDSS shows a bump in the power spectrum  at nearly the same value of
$k$ (Tegmark et al. 2003a). Such a bump would be a deviation from the
\lcdm model and would be indicative of something very interesting
happening at large scales. We have considered the possibility that
there is such a bump in the power spectrum, and we investigated if the
high level of filamentarity observed in the LCRS is indicative of
excess power at $k=0.05 \, h \, {\rm Mpc}^{-1}$  in the power 
spectrum.    

To present a brief outline of our paper, in Section 2. we present the
method of our analysis, in Section 3. we present our results and
finally in Section 4. we discuss our results and present conclusions. 

\section{Method of analysis.}
The LCRS has six slices,  each $1.5^o$ thick in declination and $80^o$
wide in right ascension. Three of the slices are in the Northern
galactic cap region centered around  declinations $-3^o$, $-6^o$
and $-12^o$,   and the other three are in the Southern galactic cap
region at declinations $-39^o$, $-42^o$ and $-45^o$. We extracted
volume limited subsamples with absolute magnitude
limits $-21.5 \le M \le -20.5$ so as to uniformly
sample  the region  from $195$ to $375 \, h^{-1}$Mpc in the radial
direction. The subsamples used here and the method of analysis are
exactly the same as in Bharadwaj et al. (2004).      

Our data consist of a total of 5073 galaxies distributed in 6 slices.  
The slices were all collapsed along  the thickness (in declination)
resulting in a 2 dimensional truncated conical slice. Each slice was 
unrolled and then embedded in a $1 \, h^{-1} {\rm Mpc} \times 1 \, 
h^{-1} {\rm  Mpc}$  2D rectangular grid.  Grid cells with galaxies in
them were assigned the value 1, empty cells 0. Connected regions of
filled cells were  identified as clusters using  a
``friends-of-friends''(FOF) algorithm. The geometry and topology of
a two dimensional cluster can be described by the three Minkowski
functionals, namely its area $S$, perimeter $P$, and genus $G$. It is
possible to quantify the shape of the cluster using a single  2D
``Shapefinder'' statistic  (Bharadwaj et al. 2000) which is defined 
as the dimensionless ratio
\begin{equation}
{\cal F}=\frac{P^2 - 4 \pi S}{P^2 + 4 \pi S}\,, 
\end{equation}
which by construction  has values in the range $0 \le {\cal F} \le 1$. It
  can be verified that  ${\cal 
  F} =1$ for an ideal filament which has a finite length and zero
  width, whereby it subtends no area ($S=0$) but has a finite
  perimeter ($P>0$). It can be further checked that ${\cal F}=0$ for a
  circular disk, and intermediate values of ${\cal F}$ quantifies the
  degree of filamentarity with the value increasing as a cluster is
  deformed from a circular disk to a thin filament. 

The definition of ${\cal F}$ needs to be modified when working on a
rectangular  grid of spacing $l$.  An ideal filament, represented on a
grid,   has the minimum possible width {\it i.e.} $l$,  and its perimeter
$P$ and area $S$ are related as 
$P=2 S + 2 l$. At the other extreme we have  $P^2=16 S$ for a square 
shaped cluster on the grid. We introduce the  2D Shapefinder statistic 

\begin{equation}
{\cal F} = \frac{(P^2 - 16 S)}{(P-4 l)^2}
\end{equation}
to quantify the shape of  clusters on a grid.  By definition 0$\le
{\cal F} \le$ 1.  ${\cal F}$ quantifies the degree of filamentarity of the 
cluster, with ${\cal F}$ = 1 indicating a filament  and ${\cal F}$ =
0, a square, and ${\cal F}$ changes from $0$ to $1$ as a square is
deformed to a filament.  

Instead of studying the filamentarity of individual clusters, we use
the average filamentarity to asses the morphological properties of the
overall galaxy distribution.   The average filamentarity $F_2$ is
defined as the mean filamentarity  of all the clusters in a slice
weighted by the square of the area of each  clusters 
\b F_2 = {\sum_{i} {\cal S}_i^2 {\cal F}_i\over
  \sum_{i}{\cal S}_i^2} \,. \e 
The extent of filamentarity in a survey is strongly influenced by the
morphology of its largest and most massive members. A supercluster
will contribute more to the overall texture of large-scale structure
than an individual cluster of galaxies, and we incorporate this by
using the second area-weighted moment of the filamentarity.   
In the current analysis, we use the average filamentarity to quantify
the degree of filamentarity in each of the LCRS slices.

The galaxy distribution in the LCRS slices is quite sparse,  and
the Filling Factor $FF$, defined as the fraction of filled cells,  is
very small ($FF \sim 0.01$). The clusters identified using  FOF
contain at most 2 or 3 filled cells, and these do not  correspond  to
the long filaments visible in the LCRS slices.  It is necessary to
coarse-grain, or smoothen,  the galaxy distribution before it is
possible to objectively identify the large-scale coherent structures
picked out by our eyes. The coarse-graining  is achieved by 
successively filling cells that are immediate neighbors of already
filled cells. The filled cells get fatter after every iteration of 
coarse-graining. This causes clusters to grow, first because of the
growth of filled cells, and then by the merger of adjacent clusters as
they overlap. At the initial stages of coarse-graining,  the patterns
which emerge from the distribution of 1s and 0s closely resembles the
coherent  structures  seen in the galaxy distribution. As the
coarse-graining proceeds, the clusters become 
very thick and fill up the entire region washing away any patterns.
$FF$ increases from $FF \sim$ 0.01 to $FF =$ 1 as the coarse graining
proceeds.  So as not to limit ourselves to an arbitrarily chosen value
of $FF$, we present  our results showing the average filamentarity
$F_2$ for the entire range of filling factor $FF$ (eg. Figure 2). 
We use the plots of average filamentarity as a function of the filling
factor to asses the overall morphology of the large-scale
structures at various levels of smoothing.   It may also be noted that
the length-scale associated with the structures increases with the
filling factor, and we have a single connected structure which
percolates through the entire survey region as $FF \rightarrow 1$.

The cosmological simulations were carried out using a Particle Mesh
(PM) N-body code. A comoving volume $[409.6  h^{-1} {\rm Mpc}]^3$ was
simulated using $256^3$ particles on a $512^3$ mesh. The set of values 
$(\Omega_{m0},\Omega_{\Lambda0},h)=(0.3,0.7,0.7)$ were used for the
cosmological parameters, and we used  a $\Lambda {\rm CDM}$ power
spectrum characterized by a spectral index $n_s=1$ at large-scales and
with a value $\Gamma=0.2$  for the shape parameter.  The power
spectrum was normalized to $\sigma_8=1.0$, consistent with the COBE 
data,  which gives a good fit to the LCRS power spectrum.  We have also 
run simulations using $\sigma_8=0.84$ which is more in keeping with the
WMAP results. We find that the results for the filamentarity are not
very sensitive to the normalization.  There is practically no
difference if we use any one of the two above values, so we report 
results only for $\sigma_8=1.0$. 

The N-body simulation gives us the final distribution of dark
matter particles for different random realizations of the initial
density fluctuations. We have run the simulation for three different
realizations of the initial density fluctuations. In order to extract
galaxies corresponding to the LCRS, the observer was placed at a
suitable position in the N-body simulation cube  and the dark matter
peculiar velocities were used to take whole distribution  over to
redshift space. For each volume limited LCRS slice we identified a
region in the N-body simulation which has exactly the same shape and size
as the LCRS slice that  we analyzed, and we extracted exactly the same
number of dark matter particles  as there  are galaxies   in the
slice. For each LCRS slice we extracted three independent simulated
slices  located at three different orientations within a single N-body
cube. This,  combined with the three independent realizations of the
initial density fluctuations,  gives  us nine   independent
simulations of each LCRS slice.  These simulated slices, extracted 
from the N-body simulation were analyzed in exactly the same manner as
the actual LCRS data.  

Various lines of investigation indicate that galaxies may be a biased
tracer of the underlying dark-matter distribution (eg. Peacock \&Dodds
1994). Theoretical considerations (eg. Kauffmann et al. 1996, Scherrer
\& Weinberg 1998) also suggest that constant linear biasing  holds on
the large scales of interest here.  We adopted the ``sharp cutoff''
scheme (Cole et al. 1998),  a local  biasing scheme where  the
probability of a mass particle being selected as a galaxy is a function
of the neighbouring density field alone. In this scheme the  final
dark-matter distribution generated by the N-body simulation was first 
smoothened  with a Gaussian of width 5 $h^{-1}$ Mpc. Only particles
which lie in regions where the density contrast, in units of the
rms. density fluctuation, exceeds a critical value were
chosen. Varying the value of the critical density contrast changes the
bias. We chose the values for the critical density contrast so as to
produce particle distributions with a modest bias $b=1.15$ and
high bias $b=1.5$. The simulated LCRS slices were extracted from the
biased particle distribution in exactly the same manner as described
for the unbiased case.   

We have run three independent realizations of the N-body simulation
for a model where the power spectrum has excess power at scales $k
\sim 0.05 \, h \, {\rm Mpc}^{-1}$. In these simulations the power at
scales between $0.04 \, h \, {\rm Mpc}^{-1}$ to $0.06 \, h \, {\rm 
  Mpc}^{-1}$ was increased by  $100 \%$, thereby introducing a bump
in the power spectrum centered at $k =  0.05 \, h \, {\rm Mpc}^{-1}$.   
The simulated LCRS slices were extracted in exactly the same way as
described earlier. 

\section{Results}
We first show, in Figure 1, the galaxy distribution in one
of the LCRS slices ($\delta=-3^{\circ}$) along with simulated slices
for the \lcdm model, the \lcdm model with a high bias ($b=1.5$) and
the \lcdm model with a bump. Comparing the simulated  \lcdm slices  
with and without the bump shows that the bump does not seem to make any
difference in the coherent structures, at least as far as the eye can
make out. Introducing bias, we find, makes a difference. In the biased
slice the galaxies are preferentially  chosen from  the  dense regions
which appear to be much more pronounced. Most of the galaxies being 
located in the very dense regions,  the voids are
larger  and the biased slice  seems to lack some of the very large
scale coherent structures seen in the unbiased \lcdm slice. Visually
comparing the actual LCRS with 
the simulated slices shows that \lcdm seems to reproduce the
large-scale coherent features reasonably well. The galaxy distribution
in the simulated \lcdm slice appears to be a little more diffuse in
comparison to the actual slice, possibly indicating  a small amount of
bias.

\begin{figure}
\figurenum{1}
\epsscale{0.7}
\rotatebox{-90}{\plotone{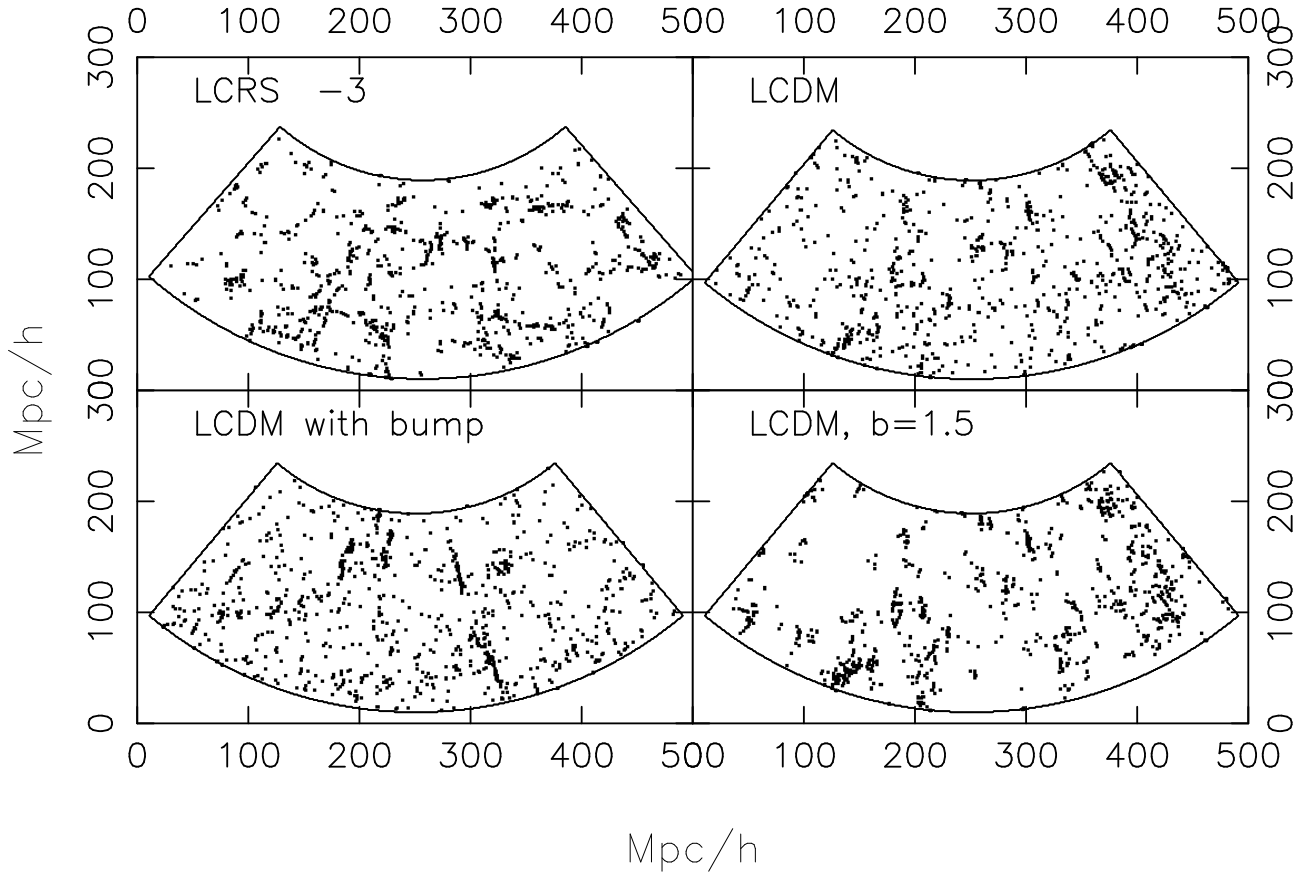}}
\caption{This shows the galaxy distribution in the LCRS slice at
  declination $\delta=-3^{\circ}$, along with the simulated slices as 
  indicated in the panels. These figures show the galaxy distribution
  after it has been converted to a set of 1s and 0s on a grid, and
  after we have applied two iterations of the coarse-graining
  procedure described in Section 2. This, to some extent, enhances for
  our eyes  the large scale coherent structures present in the galaxy 
  distribution.}
\end{figure}

The results of our quantitative analysis of filamentarity using
Shapefinders are shown in Figures 2 and 3. For each simulated slice we
have used  nine independent realizations to estimate  the mean  and 
the variance for the average filamentarity $F_2$ as a function of the 
filling factor $FF$. Earlier investigations (Bharadwaj et al. 2004)
showed that the large-scale features are washed out at the level of 
coarse-graining where   $FF \sim 0.7$, so we restricted
our analysis to the range  $FF \le 0.7$ only. We used  $\chi^2/\nu$  
 the  reduced $\chi^2$ per degree of freedom  (Table I) to quantify
 the deviations in the average filamentarity 
between the models and the actual LCRS slices. 

We find that the average filamentarity in  the unbiased \lcdm model 
is somewhat lower than the actual values for most of the slices
(Figure 2). Introducing a small bias $(b=1.15)$ increases the average
filamentarity at low filling factors, leaving the values at large
filling factors  unaffected. The fit to the actual data, as
quantified by the values of $\chi^2/\nu$, shows a considerable
improvement. Two of the slices ($\delta=-3^{\circ}, -45^{\circ}$) have
values of $\chi^2/\nu$ in the range $1.7$ to $1.8$, and  it is less
than $1.1$ for all the other slices. Given the lack of knowledge about
the statistical properties of the errors in $F_2$, we may consider
these values of $\chi^2/\nu$ as indicative of a reasonable agreement
between this model (\lcdm,$ b=1.15$) and the actual data. A large bias
($b=1.5$) changes the  overall shape of the average filamentarity
versus filling factor curves. With increasing bias,
the average filamentarity goes up at small filling factors and 
it falls at large filling factors. Relating the filling factor to the
size of the structures, we find that bias enhances filamentarity on
small scales and suppresses the large-scale filamentary patterns. This
is in keeping with the visual impression discussed earlier. 

\begin{figure}
\figurenum{2}
\epsscale{0.7}
\rotatebox{-90}{\plotone{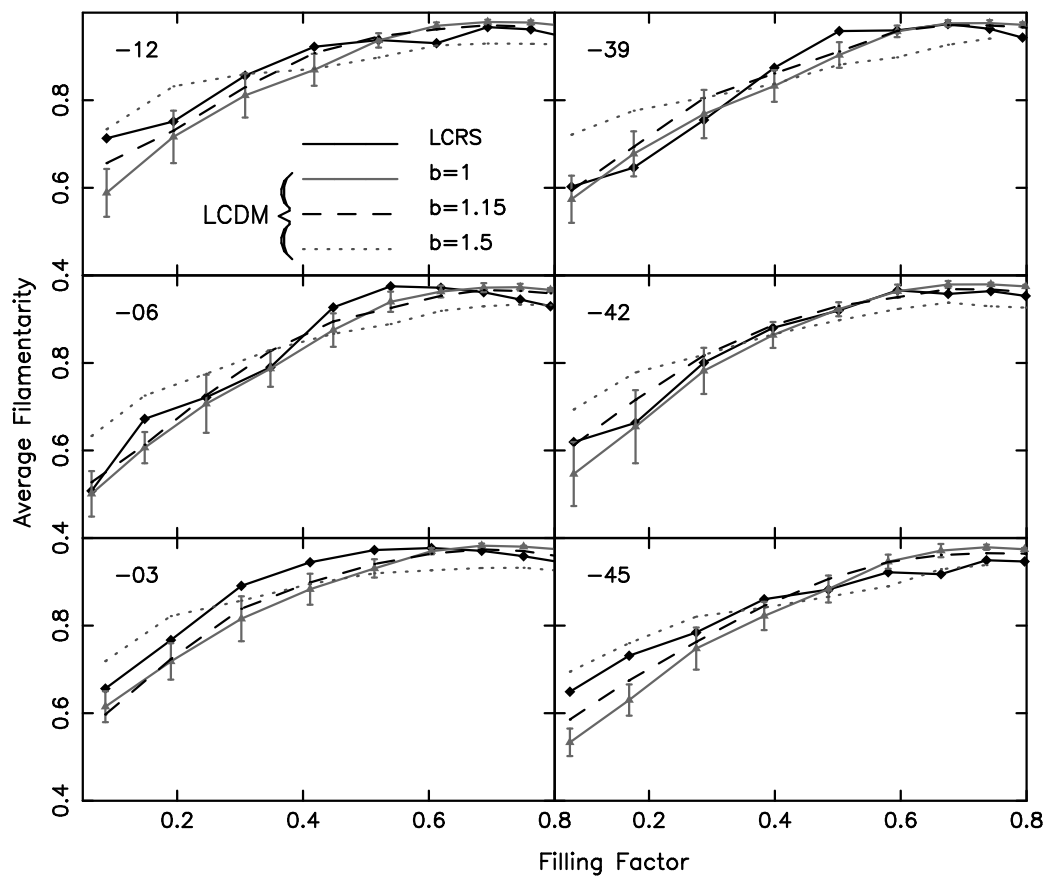}}
\caption{This shows the average filamentarity as a function of the
  filling factor for the six LCRS slices along with the results for
  the slices simulated in the \lcdm model with the different values of
  bias shown in the figure. Nine independent realizations of  each of
  the  simulated slices  were used to calculate the mean and variance
  of the average filamentarity. The $1-\sigma$ error-bars are shown
  only for the unbiased \lcdm, the error-bars are similar for the
  biased models.} 
\end{figure}

The results for the simulations using a \lcdm model with a bump are
shown in Figure 3. Introducing a bump in the power spectrum does not
seem to make a noticeable change in the average filamentarity. This
is further reinforced by the values of $\chi^2/\nu$ which are very
similar to the values when there is no bump.  

\begin{figure}
\figurenum{3}
\epsscale{0.7}
\rotatebox{-90}{\plotone{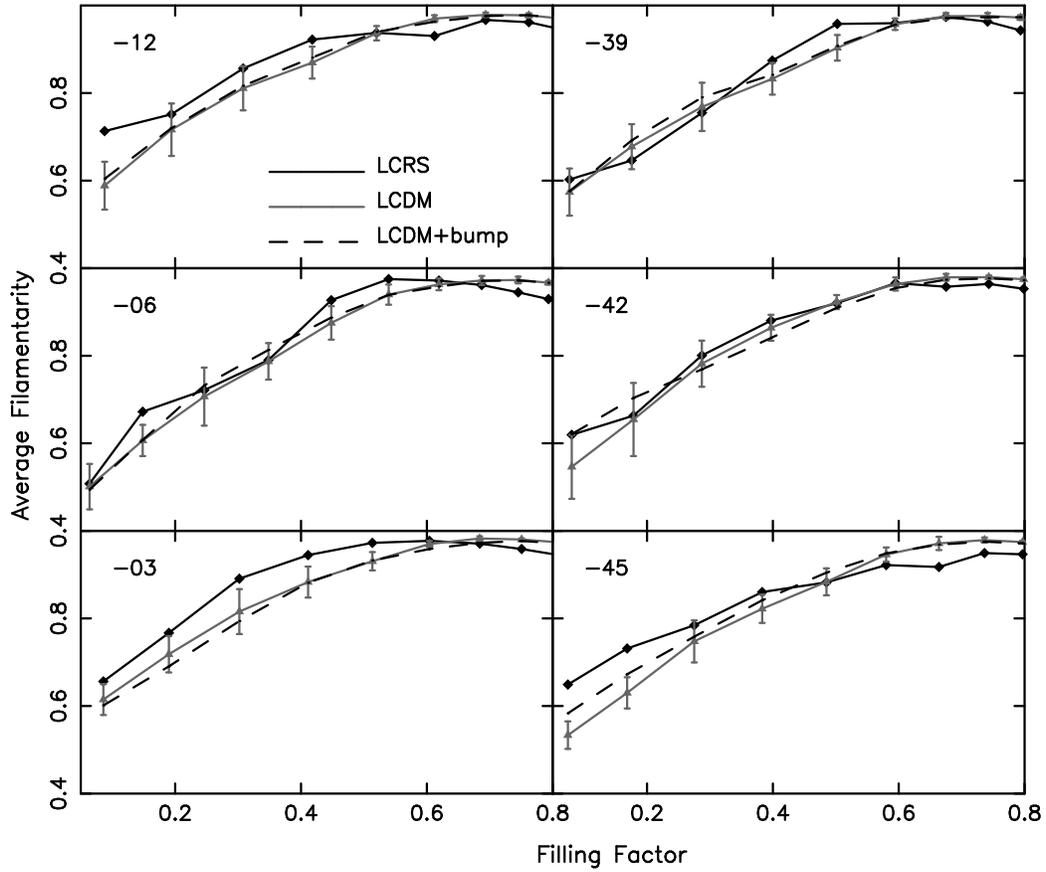}}
\caption{This shows the average filamentarity as a function of the
  filling factor for the six LCRS slices along with the results for
  the slices simulated in the \lcdm model with and without the
  bump. The $1-\sigma$ error-bars, shown only for the case without a
  bump, have similar values when a bump is introduced.}
\end{figure}

\begin{center}
\begin{table}
\caption{For the LCRS slices at different declinations $\delta$,  we
  tabulate below  the values of  $\chi^2/\nu$  for the \lcdm model
  with different values of bias and with a bump in the power
  spectrum.} 
\begin{tabular}{c|cccc}
\hline
$\delta$    & $b=1$ & $b=1.15$ & $b=1.5$  & Bump \\
\hline 
$-3$   & 2.7  & 1.7  & 3.0 & 2.9 \\
$-6$   & 1.8  & 1.1  & 4.3 & 1.2 \\
$-12$  & 5.7  & 1.1  & 1.5 & 3.8 \\
$-39$  & 0.89 & 0.80 & 5.7 & 0.77\\
$-42$  & 1.26 & 0.42 & 1.2 & 0.50  \\
$-45$  & 5.5  & 1.8  & 1.0 &  2.7 \\
\hline
\end{tabular} 
\end{table}
\end{center}
\section{Discussion and Conclusions}
The \lcdm model has been found to be consistent with  very precise
observations of large-scale structures in the universe and their
imprint on the CMBR. These tests are limited by the fact that  most of
them are based on  two-point statistics  which are not sensitive
to the large-scale coherent features like the long filamentary
patterns seen in galaxy redshift surveys.  It has  recently been 
established, for the LCRS, that the filaments are statistically
significant to scale as large as $70$ to $80 \, h^{-1} \, {\rm Mpc}$ 
(Bharadwaj et al 2004).  In this paper we have tested whether the
\lcdm model  is consistent with the high level of filamentarity
observed in the LCRS. A point to note is that the analysis presented
here is restricted to two dimensional sections, and structures which
appear as filaments could actually be sections of three dimensional
planar structures. This issue whether the 2D filaments are actually
also filaments in 3D, though not of crucial importance in the present
discussion, is an interesting question which we plan to address in 
later work.  

The filamentarity in the unbiased \lcdm model is lower than in the
LCRS. Introducing a positive bias increases the 
filamentarity at small scales and suppresses the filamentarity at
large scales. As the bias is increased, the galaxy distribution gets
more concentrated in the high density regions, and the very large
scale structures are suppressed. The values of the average
filamentarity as a function of the filling factor (Figure 2)  are 
sensitive to the bias, and the quantitative analysis of  filamentarity
holds the possibility of being a sensitive probe of the  bias
parameter. We have used two values of the bias parameter and find that
a large bias $(b=1.5)$ is ruled out. The simulations with a  smaller
value of bias $(b=1.15)$ give a good fit to most of the  LCRS slices,
though in two of the slices ($\delta=-3^{\circ}, \, -45^{\circ}$) the
average filamentarity in the simulations falls somewhat below the LCRS
for nearly the entire range of filling factor. The statistical 
significance of this mild discrepancy ($\chi^2/\nu \sim 1.8$) is not 
straightforward to interpret, and we adopt the point of view that the
\lcdm model with a small bias is consistent with the filamentarity in
the LCRS slices. It is interesting to note that the LCRS
$\delta=-3^{\circ}$ slice, where the filamentarity is somewhat in
excess of the biased \lcdm model, also has statistically significant
filaments extending to scales ($70 - 80 \, h^{-1} \, {\rm
  Mpc}$) larger compared to the other slices  ($50 - 70 \, h^{-1} \, {\rm
  Mpc}$) (Bharadwaj et al 2004).   Analysis of the filamentarity in
larger redshift  surveys and a better understanding of the statistical
properties of the errors in the average filamentarity will, in the
future, allow this test to impose more stringent constraints on models
of structure formation. A point to note is that the values of bias
$b=1.15$ and $1.5$ have no special significance,  and have been
chosen as two convenient values one representing a modest bias and
another a high bias. The value $b=1.15$ is consistent with the
analysis of the LCRS power spectrum  (Lin et al. 1996) who conclude
that the value of the bias in the LCRS is in the range $0.7$ to
$1.3$. A point to keep in mind is that the  volume limited subsamples
analyzed here contain only the very  bright galaxies, and the bias is
known to increase with the intrinsic luminosity of the galaxies
(eg. Wild et al. 2004)  

There have been speculations that there may be a bump in the power
spectrum around the wave number $k \sim 0.05 \, h \, {\rm Mpc^{-1}}$,  and
this may have an influence on the filamentary pattern seen in galaxy
redshift surveys. We have considered a bump in the power spectrum
which enhances the power in the range $0.04$ to  $0.06 \, h \, {\rm
  Mpc^{-1}}$, and we find that this does not have a significant influence
on the average filamentarity at any value of the filling factor. The
bump in the power spectrum, if it exists, seems to be unrelated to the
filamentary patterns seen in redshift surveys. 
  
It is interesting to compare our results with some of the other tests
which probe models of structure formation beyond the two-point
statistics. The bispectrum goes one step beyond the power spectrum, and
is sensitive to non-Gaussian features. It has been used to test
non-Gaussianity in the primordial power spectrum and determine the
bias parameter (eg. Verde at al. 2002, Scoccimarro et al. 2001), but
it does not tell us very much about individual, coherent features like
filaments. The genus statistics quantifies the topology of the galaxy
distribution. Studies of the 2D genus curve for the 2dFGRS (Hoyle
et. al 2002a) and the SDSS  (Hoyle et. al 2002b) are consistent with
the predictions of the \lcdm model. Schmalzing and Diaferio (2000)
have calculated the Minkowski functionals of the galaxy distribution
in the nearby universe (the Updated Zwicky Catalogue) and they compare
this with \lcdm N-body simulations. They find that galaxy distribution
in the simulated distributions is less coherent than what is
observed. Sheth et al. (2003) have developed a method called
``Surfgen'' for generating triangulated surfaces from a discrete
galaxy distribution, and have used this to calculate the 3D
Shapefinders. They have applied this to a variety of simulated data
(Shandarin, Sheth and Sahni 2003, Sheth 2003), but results are awaited
from real redshift surveys.  

In conclusion we note the large-scale coherent features seen in galaxy
redshift surveys provides a unique testing ground for probing the
models of structure formation beyond the two-point statistics. These
tests are also sensitive to the bias and hold the possibility of
giving accurate estimates of the bias parameter. The analysis reported
in this paper shows the \lcdm model with a modest bias to be roughly 
consistent with the filaments observed in the LCRS. Future work using
large redshift surveys like the 2dFGRS and the SDSS should be able to
accurately  quantify the network of  large scale coherent
structures and  place stringent limits on models of structure
formation, taking precision cosmology into new grounds beyond the realm
of two point statistics.  

\section{Acknowledgment}
The authors wish to thank the LCRS team for making the survey data
public. SB would like to acknowledge financial support from the Govt.
of India, Department of Science and Technology (SP/S2/K-05/2001). BP
would like to thank the CSIR, Govt. of India for financial support
through a JRF fellowship.

\end{document}